# Silicon graphene Bragg gratings

**José Capmany[1],* David Domenech[2], and Pascual Muñoz[1]**

[1]*ITEAM Research Institute, Universitat Politécnica de Valencia, Camino de Vera s/n, 46022 Valencia, Spain*
[2]*VLC Photonics S.L., Camino de Vera s/n, 46022 Valencia, Spain*
* *jcapmany@iteam.upv.es*

**Abstract:** We propose the use of interleaved graphene sections on top of a silicon waveguide to implement tunable Bragg gratings. The filter central wavelength and bandwidth can be controlled changing the chemical potential of the graphene sections. Apodization techniques are also presented.

## 1. Introduction

Graphene is a two dimensional single layer of carbon atoms arranged in an hexagonal (honeycomb) lattice featuring an energy vs momentum dispersion diagram where the conduction and valence bands meet at single points (Dirac points) [1-3]. In the vicinity of a Dirac point, the band dispersion is linear and electrons behave as fermions with zero mass, propagating at a speed of around $10^6$ m $s^{-1}$ and featuring mobility values of up to $10^6$ $cm^2V^{-1}s^{-1}$. The density of states of carriers near the Dirac point is low, and as a consequence, its Fermi energy can be tuned significantly with relatively low electrical energy (applied voltage) [1]-[4]. This tuning modifies the refractive index of graphene and, if this material is incorporated into integrated dielectric waveguide structures, it also affects the effective index and the absorption of the propagated modes opening new possibilities of obtaining tunable components in different regions of the electromagnetic spectrum. Devices exploiting this effect have been reported, for instance, in the microwave [5], terahertz [3] and photonic regions [4] of the electromagnetic spectrum.

A particularly active area of research during the last years is related to the design of tunable integrated photonic components where different contributions have theoretically and experimentally reported a variety of functionalities including electroabsorption modulation in straight waveguides [5]-[8] and resonant structures [9], channel switching [10], and electrorefractive modulation [11]. Most of these are based either on a straight waveguide configuration or in ring cavities. In this paper we propose the use of interleaved graphene sections on top of a silicon waveguide to implement tunable distributed Bragg gratings. Some properties of graphene relevant to its tunable conductivity and dielectric constant are briefly reviewed in section 2. The proposed silicon graphene waveguide Bragg grating (SGWBG) structure is presented in section 3. We begin describing the configuration of the tunable silicon graphene upon which it is based and then illustrate by numerical means the variation of the effective indexes of the transversal magnetic (TM) and electric (TEM) fundamental modes by suitable change of the chemical potential. The principle of the SGWBG is then introduced, which is based on interleaving sections of graphene on top of a silicon waveguide. In section 4 we explore the behavior of the proposed device including its wavelength tunability, bandwidth and also the possibility of side-lobe reduction by apodization of the chemical potential. Section 5 concludes the paper with a summary and some relevant future directions of research

## 2. Graphene conductivity and dielectric constant

Graphene has noteworthy optical properties due to its conical band structure that allow both intra-band and inter-band transitions [1]-[3]. Both types of transitions contribute to the material conductivity [12].

$$\sigma(\omega,\mu_c)=\sigma_{\text{intra}}(\omega,\mu_c)+\sigma_{\text{inter}}(\omega,\mu_c) \quad (1)$$

Where:

$$\sigma_{\text{intra}}(\omega,\mu_c)=\frac{ie^2}{\pi\hbar^2(\omega+i2\Gamma)}\left[\frac{\mu_c}{k_BT}+2\ln\left(e^{-(\mu_c/k_BT)}+1\right)\right] \quad (2)$$

and, if $k_BT << |\mu_c|, \hbar\omega$:

$$\sigma_{\text{inter}}(\omega,\mu_c)\approx\frac{-ie^2}{4\pi\hbar}\ln\left(\frac{2|\mu_c|-(\omega-2i\Gamma)\hbar}{2|\mu_c|+(\omega-2i\Gamma)\hbar}\right) \quad (3)$$

In the above expressions $e$ represents the charge of the electron, $\hbar$ the angular Planck constant, $k_B$ the Boltzman constant, $T$ the temperature, $\mu_c$ is the Fermi level or chemical potential and:

$$\Gamma=\frac{e\mathbf{v}_\mathbf{F}^2}{\mu\mu_c} \quad (4)$$

is the electron collision rate which is a function of the electron mobility $\mu$ and the Fermi velocity in graphene $\mathbf{v}_\mathbf{F}\approx 10^6 ms^{-1}$. The dielectric constant of graphene follows from (1)-(4):

$$\varepsilon_g(\omega,\mu_c)=1+\frac{i\sigma(\omega,\mu_c)}{\omega\varepsilon_o\Delta} \quad (5)$$

Where $\Delta=0.34\ nm$ is the thickness of the layer. Tunability of the dielectric constant is achieved by suitable application of a voltage $V_g$ to the graphene layer, since this changes the value of the chemical potential according to [11]:

$$|\mu_c(V_g)|=\hbar\mathbf{v}_\mathbf{F}\sqrt{\pi|\eta(V_g-V_o)|} \quad (6)$$

Where $Vo=0.8$ volt is the offset from zero caused by natural doping and $\eta=9\text{x}10^{16}V^{-1}m^{-2}$ [6].

### 3 Graphene silicon waveguide and Bragg grating

Graphene can be incorporated into silicon to implement graphene silicon waveguides (GSWs). For the implementation of the Bragg grating we consider the layout shown in figure 1 that consists in placing a monolayer graphene sheet on top of a silicon bus waveguide, separated from it by a thin $Al_2O_3$ layer.

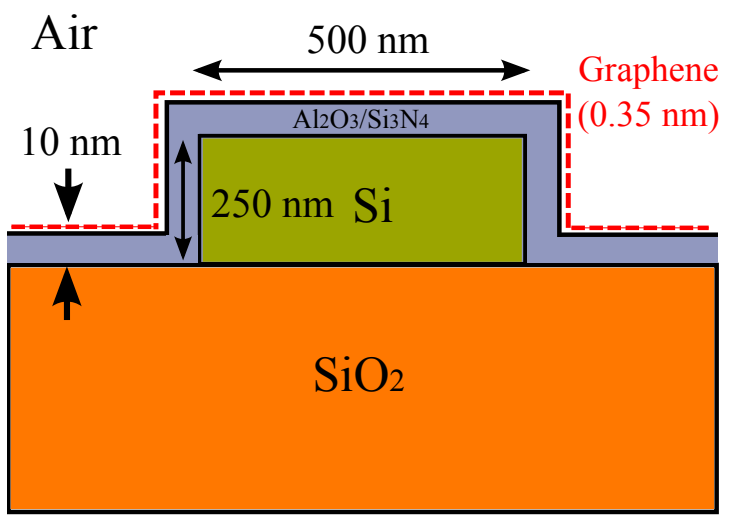

Fig. 1. Deep silicon waveguide with a layer of graphene placed on top of it.

The incorporation of the graphene layer modifies the propagation characteristics (field profile, losses and effective index) of the guided modes and these can be, in turn, controlled and reconfigured changing the chemical potential by means of applying a suitable voltage. Furthermore, these properties are wavelength dependent so a complete description requires the use of numerical and or mode solving techniques. In figure 2 we show for λ=1550 nm, the effective index and the losses (in insets) vs the chemical potential for the transversal magnetic (TM) fundamental mode (a similar behavior is obtained for the transversal electric TE mode). These results have been numerically calculated using a Finite Difference (FD) based commercial *Field Designer* mode solver, from *PhoeniX Software B.V*.

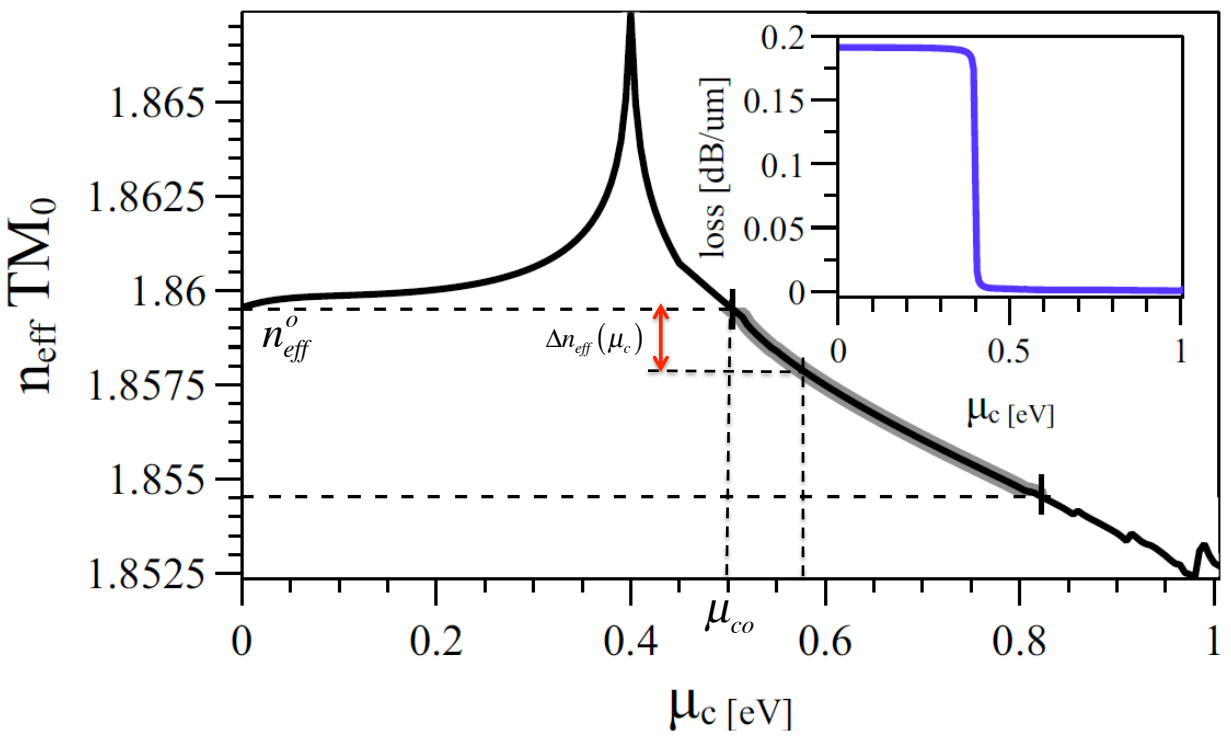

Fig. 2. Effective index and losses (inset) for the TM fundamental mode of a deep GSW versus the chemical potential.

Referring to figure 2, the value of effective index of the TM mode when no graphene is placed on top of the waveguide ($n_{eff}^{o} = 1.8595$) corresponds to the case of $\mu_c=0$ (with no losses). Note that this same value is obtained for $\mu_c=\mu_{co}$ close to 0.52 eV. Note as well that increasing the value from 0.52 to 0.82 eV results in a tunable change of the effective index $\Delta n_{eff}(\mu_c) = \left| n_{eff}(\mu_c) - n_{eff}^{o} \right|$ in the range of 0 to 0.3%. This effect can be exploited to implement a tunable Bragg grating by interleaving graphene sections on top of the silicon waveguide as shown in figure 3.

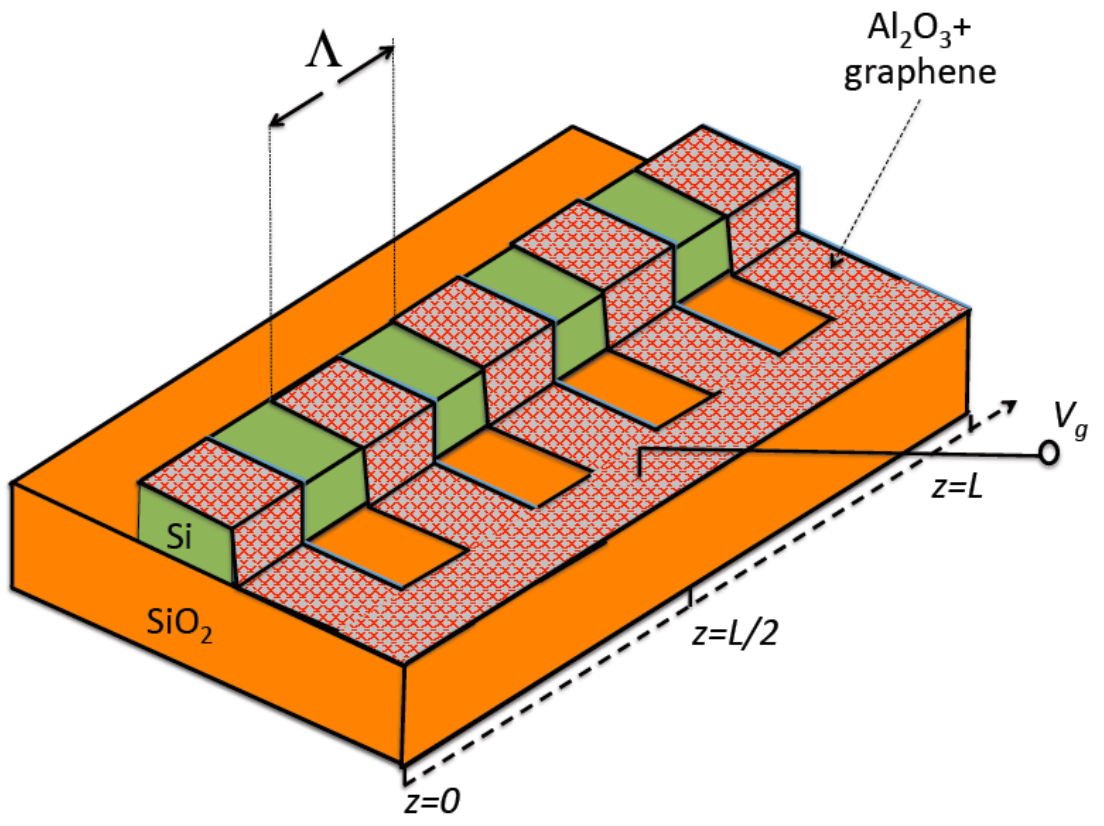

Fig. 3. Layout of the proposed silicon graphene Bragg grating.

## 4 Results and discussion

For a grating structure of length $L$ and period $\Lambda$ the *N-th* order resonance central operating wavelength $\lambda_D$ and grating bandwidth are given by [13]:

$$\lambda_D(\mu_c, N) = 2\bar{n}_{eff}(\mu_c)\Lambda / N$$
$$\bar{n}_{eff}(\mu_c) = \frac{n^o_{eff} + n_{eff}(\mu_c)}{2} \tag{7}$$

$$\frac{\Delta\lambda(\mu_c, N)}{\lambda_D(\mu_c, N)} = \frac{\Delta n_{eff}(\mu_c, N)}{\bar{n}_{eff}(\mu_c)} \sqrt{1 + \left(\frac{\lambda_D(\mu_c, N)}{\Delta n_{eff}(\mu_c, N) L}\right)^2} \tag{8}$$

All of them can be modified changing the applied voltage $V_g$ (and the chemical potential) to the interleaved graphene sections.

A Bragg grating has been designed with the following parameters: N=3, Λ=1250.4 nm *L=1500 μm, μ<sub></sub>* $\mu_{co}$*=0.52 eV*, $n^o_{eff} = 1.8591$. Figure 4 shows the results for the power reflection and transmission transfer functions of the device versus the optical wavelength, taking the chemical potential as a parameter. The results were computed by numerical solution of the coupled mode equations [13].

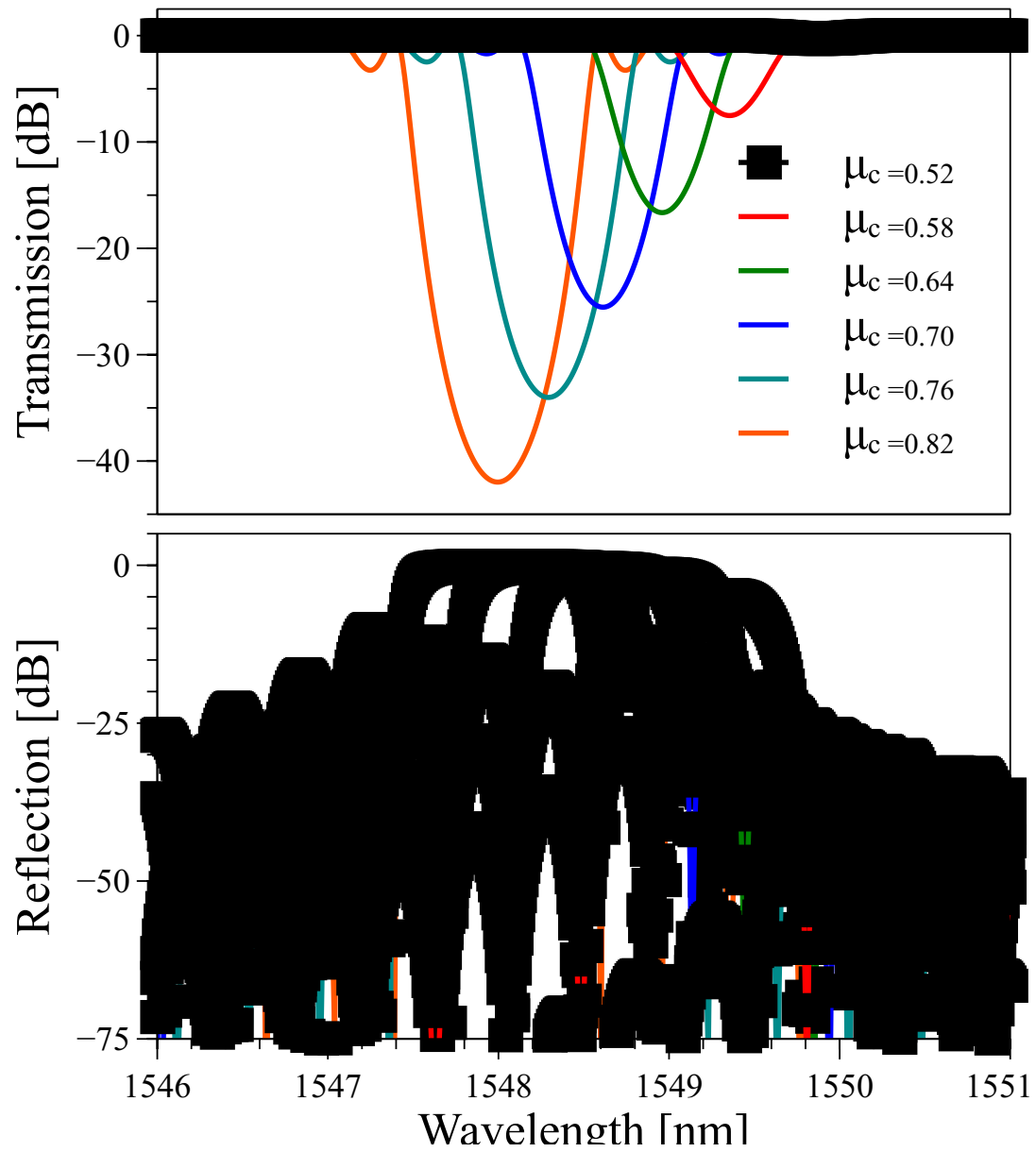

Fig. 4. Transmission (Upper) and Reflection (Lower) intensity transfer fucntion of a Silicon graphene Bragg grating with *N=3 Λ=1250.4 nm, L=1500 μm,* $\mu_{co}$*=0.52 eV*, $n_{eff}^o$=1.8591.

As it can be observed as the chemical potential is increased the grating central wavelength is shifted to lower values. Note as well that the grating bandwidth and the maximum value of the reflection increase. These features can be explained from Equations (7), (8) and the dependence of the effective index on the chemical potential shown in figure 2. Note from figure 2 that as $\mu_c$ increases then $n_{eff}(\mu_c)$ decreases and hence $\bar{n}_{eff}(\mu_c)$ and $\lambda_D(\mu_c)$ decrease as well. The non-constant value of the average effective index with $\mu_c$ is due to the fact that while the change in the effective index is increased with $\mu_c$ the maximum value remains constant. On the other hand $\Delta n_{eff}(\mu_c)$ increases with $\mu_c$ and the quotient $\Delta n_{eff}(\mu_c)/\bar{n}^o_{eff}(\mu_c)$ dominates over the square root factor in equation (8). Figure 5 shows the evolution of the central operating wavelength $\lambda_D$ and the grating bandwidth obtained by numerical simulation.

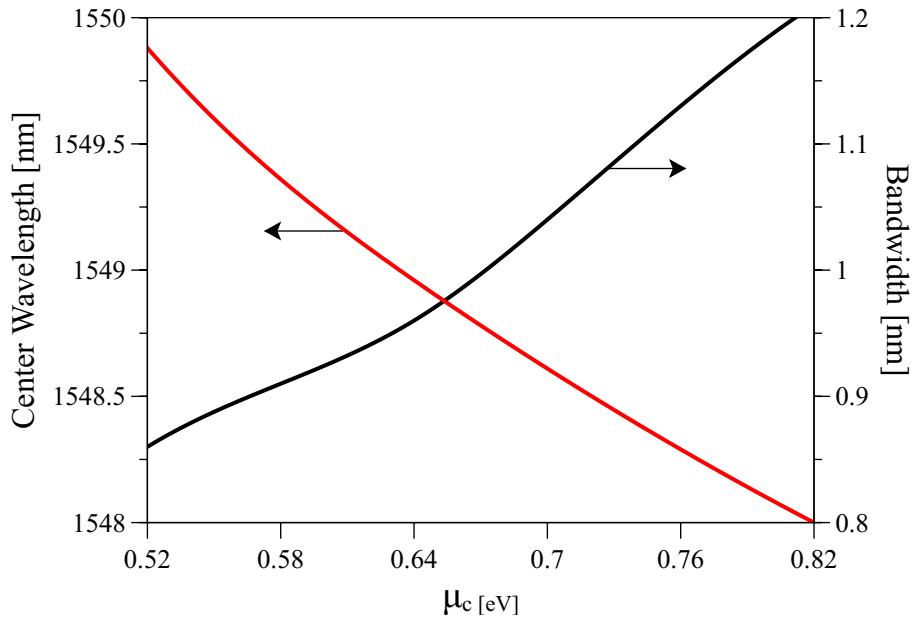

Fig. 5. Central operating wavelength $\lambda_D$ and the grating bandwidth evolution versus the chemical potential of a Silicon graphene Bragg grating with *Λ=1250.4 nm, L=1500 µm,* $\mu_{co}$*=0.52 eV*, $n_{eff}^o$=1.8591.

The proposed design can also implement apodized Bragg grating configurations which are useful to modify their spectral response in order to reduce the sidelobes at one or both sides of the central wavelength. This can be achieved, for instance, by changing the applied voltage in the propagation region. Figure 6 shows as an example the results for the reflection and transmission functions in a Gaussian apodized ($\mu_c=0.62$ *eV*) design where:

$$\Delta n_{eff}(\mu_c, z) = \Delta n_{eff}(\mu_c, L/2)\exp\left[-G^2\left(z-\frac{L}{2}\right)^2\right] \quad (9)$$

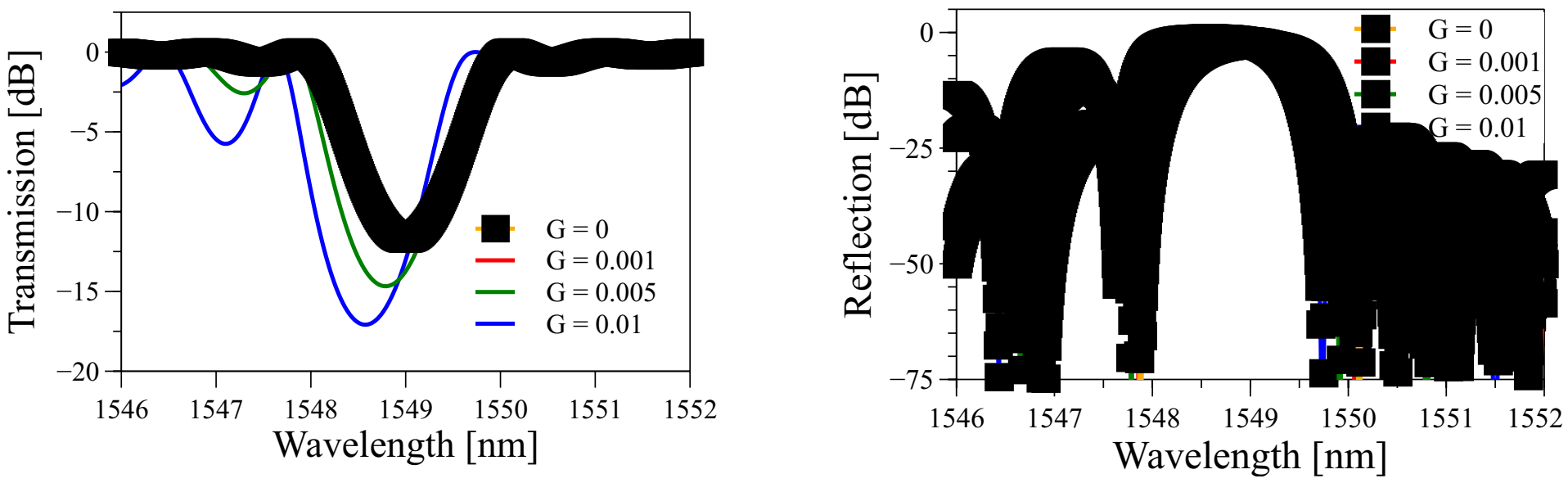

Fig. 6. Transmission (left) and Reflection (right) intensity transfer function of a Silicon graphene, Gaussian-apodized, Bragg grating with *N=3, Λ=1250.48 nm, L=1500 µm,* $\mu_c$*=0.62 eV* taking the Gaussian window coefficient G as a parameter.

In (9) $G$ is the apodization parameter that controls the steepness of the variation in refractive index as compared to the value in the central position of the grating ($z=L/2$). As expected, sidelobe suppression is obtained only in the long wavelength region taking the central wavelength of the grating as a reference. This is due to the fact that the average effective index in the grating is not constant [14]. In order to obtain a symmetric spectral behavior, several techniques such as average index pre-compensation can be implemented. These, and other techniques are under current investigation and will be reported in the near future.

### 5. Summary and Conclusions

We have proposed the use of interleaved graphene sections on top of a silicon waveguide to implement tunable Bragg gratings. The filter central wavelength and bandwidth can be controlled changing the chemical potential of the graphene sections. Apodization techniques have also been presented and discussed.

### Acknowledgments

The authors wish to acknowledge the financial support given by the Research Excellency Award Program GVA PROMETEO 2013/012.

---